\documentclass[aps,pra,twocolumn,groupedaddress,amsmath,amsfonts,amssymb]{revtex4-1}
\usepackage[english]{babel}
\usepackage{ucs}
\usepackage[utf8x]{inputenc}
\usepackage{graphicx}
\usepackage{bm}
\usepackage{bbm}
\usepackage{empheq}

\begin{document}

\title{Fidelity of optically induced single-spin rotations in semiconductor quantum dots in the presence of nuclear spins}

\author{Julia Hildmann and Guido Burkard}

\affiliation{\em
Department of Physics, University of Konstanz, D-78457 Konstanz, Germany
}

\begin{abstract}
%\noindent
We examine the influence of nuclear spins on the performance
of optically induced rotations of single electron spins in semiconductor quantum dots. 
We consider Raman type optical transitions between electron spin states
and take into account the additional effect of the Overhauser field.
We calculate average fidelities of rotations
around characteristic axes in the presence of nuclear spins analytically with perturbation theory up
to second order 
in the Overhauser field. Moreover, we calculate the fidelity using numerical averaging over the nuclear field
distribution, including arbitrary orders of the hyperfine interaction. 
\end{abstract}

\maketitle

\section{Introduction}
%\noindent
%
Single electron spins in quantum dots represent a suitable physical system for the experimental realization of
quantum bits (qubits) \cite{loss_divicenzo_proposaL_QD}.
Since single- and two-qubit operations are sufficient 
for implementing any arbitrary quantum gates \cite{divincenzo_pra1995}, a large amount
of research has been conducted for the
realization of single qubit state control and two qubit operations
\cite{burkard_prb1999, imamoglu_prl1999, petta_science2005,koppens_nature2006, cerletti_nano2005,
hanson_rmp2007, kloeffel_annrev2013}.
Along with electrical control, one of the possibilities to
manipulate single electron spin state is by optical means
\cite{imamoglu_prl1999, chen_prb2004, economou_prb2006, economou_prl2007},
which offer a fast and coherent way for control of spin state in quantum dots. \\
\indent
Experimental achievements
in optical initialization, read-out, coherent control and manipulations of single electron
spins in quantum dots
\cite{atature_science2006, mikkelsen_natphys2007, berezovsky_science2008, press_nature2008, dubin_prl2008, 
xu_natphys2008, ramsay_prl2008, loo_prb2011}
have reached a level at which their use for quantum information processing seems to be feasible.
Additionally optical control offers the possibility of incorporating electron spin qubits into hybrid
systems in which
the single spin state is entangled with the state of a photon and quantum information is
transferred by photons \cite{liu_advinphys2010, gao_nature2012, schaibley_prl2013}. \\
\indent
The accomplishment of the essential steps for optimal optical single spin control is affected by
different types of errors, e.g.
imperfection of the applied laser pulses \cite{economou_prb2006}.
Mixing of heavy and light hole states influences
the trion state, which is used 
in some schemes as an intermediate state \cite{economou_prl2007, loo_prb2011}.
However, the mixing of the heavy and light holes in quantum dots  can be controlled
e.\,g. by means of anisotropic stress \cite{plumhof_prb2013} and in this way the errors
created by the phenomena can be avoided. \\
\indent
Another intrinsic mechanism causing decoherence of electron spins in III-V semiconductor quantum dots
is their interaction with the nuclear spins
of the host material \cite{khaetskii_prl2002, coish_prb2004, coish_statsolidi2009, cerletti_nano2005}.
An electron confined in such a quantum dot interacts by hyperfine
coupling with a large nuclear bath (roughly 10$^5$-10$^6$ nuclear spins per quantum dot).
The total magnetic field of the nuclei, also called
Overhauser field, fluctuates randomly and acts as an effective magnetic field on the electron,
causing dephasing of the electron spin state.
There are possibilities to improve the decoherence time by reducing the fluctuations of the Overhauser field.
One such possibility is to polarize nuclear spins to a high degree \cite{coish_prb2004},
another is to drive or project
the nuclear spin state into an eigenstate of the Overhauser field operator
\cite{stepanenko_prl2006,issler_prl2010}.
A significant improvement of electron spin coherence time was observed in the experiments, where nuclear 
spin fluctuations were suppressed by driving the nuclear field to a stable state
\cite{xu_nature2009, bluhm_prl2010}.\\
\indent
In this paper, we focus on the single spin rotation errors arising from the interaction with an
unpolarized ensemble of nuclear spins. One possibility to rotate the single spin states
in a quantum dot is by using Raman transitions \cite{imamoglu_prl1999, chen_prb2004} between
single electron spin states split by magnetic field in Voigt geometry  via  the trion state
comprised of two electrons and a heavy hole.
In this case the transitions are driven by specifically detuned laser pulses (Fig.~\ref{level_scheme}).
The hyperfine interaction leads to a fluctuating spin state splitting and therefore
to imperfect spin rotations.\\
\indent
To compare the single spin rotations in the presence and in the absence of nuclear spins we compute
the fidelities of the unitary time evolution of the electron spin state under the action of the laser
light with and without including the hyperfine interaction.
We average the obtained fidelities over the Overhauser field distribution analytically to
second order of the hyperfine coupling and numerically to an arbitrary order.
We calculate the average fidelities for rotation axes parallel and perpendicular to the
external magnetic field and discuss the factors that influence the average rotation fidelities in both cases.\\
\indent
The previous work on the performance of the two-qubit quantum operations in the presence of the hyperfine 
interaction showed that the fidelity depends on such conditions as the gradient of the nuclear fields
between the two quantum dots \cite{hildmann_prb2011}.
The investigation of the fidelities of single-qubit rotations
exposed the dependence on the relative orientation of the rotation axis relative to the external magnetic field.
It was found that the dephasing time of the electron spin state differs if it is rotated around different axes.
It would be interesting to confirm this observation for electron spin resonance on 
single spins in electrostatically defined or semiconductor nanowire quantum dots.
\\
\indent
This paper is organized as follows. In Section II, we describe the mechanism
of Raman type optical transitions between single electron states. We include the hyperfine interaction 
to the system in Section III and derive the time evolution operator of the single-spin state in the 
presence of the Overhauser field. The calculated rotation fidelities are presented in Section IV.
\section{Optically induced single spin rotations}
%\noindent
The interaction between $\sigma_+$ polarized light in the growth 
direction $z$ and a single electron confined in a
quantum dot is given by
\begin{equation}
H=E_t\left|t_\uparrow\right\rangle\left\langle t_\uparrow\right|+
g^*(t)\left|t_\uparrow\right\rangle\left\langle \uparrow\right|+
g(t)\left| \uparrow\right\rangle\left\langle t_\uparrow \right|, 
\label{H}
\end{equation}
where $\left|\uparrow\right\rangle$ is the spin-up state in the conduction band and
$\left|t_\uparrow\right\rangle$ is the trion state formed by two electrons in the
singlet state and a heavy hole 
with angular momentum +3/2. $E_t$ is the energy of the trion state, and
$g(t)$ is the coupling to the laser field.
With an additional magnetic field applied in $x$-direction,
perpendicular to the growth direction (Voigt geometry),
the Hamiltonian (\ref{H}) reads in the basis $\left|\pm x\right\rangle=
\tfrac{1}{\sqrt{2}}(\left|\uparrow\right\rangle\pm\left|\downarrow\right\rangle), 
\left|t_\uparrow\right\rangle$ \cite{liu_advinphys2010},
\begin{equation}
 H=\begin{pmatrix}\omega_Z/2 & 0& g^*(t)/\sqrt{2}\\
 0&-\omega_Z/2 & g^*(t)/\sqrt{2}\\
g(t)/\sqrt{2} & g(t)/\sqrt{2} & E_t \end{pmatrix},
\end{equation}
where $\omega_Z$ is Zeeman splitting of the electron states $\left|\pm x\right\rangle$.
Applying two-color laser pulses enables arbitrary rotation
of the electron spin by a single pulse \cite{chen_prb2004},
$$g(t)=\Omega_1(t)e^{-i\omega_1 t-i\alpha}+\Omega_2(t)e^{-i\omega_2 t}
$$
with Raman resonance conditions:
$$\omega_1+\omega_Z/2=\omega_2-\omega_Z/2= E_t-\Delta, 
$$
\begin{figure}[t!]
\centering\includegraphics[width =0.4\textwidth]{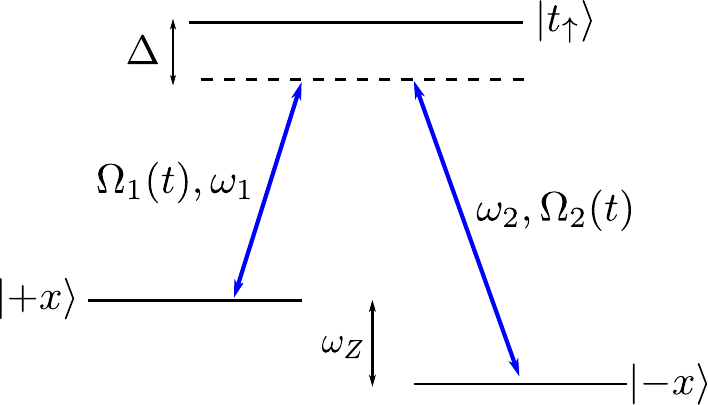}
\caption{Energy level scheme for an electron spin interacting with 
$\sigma_+$ polarized light.
The transverse magnetic field splits $x$-eigenstates of the electron spin by the Zeeman energy $\omega_Z$.
The electron spin states are virtually coupled to the trion state $|t_\uparrow\rangle$
by the laser pulses $\Omega_1(t)$
and $\Omega_2(t)$ with frequencies $\omega_1$ and $\omega_2$, with detuning $\Delta$.}
\label{level_scheme}
\end{figure}
where $\Delta$ is the laser detuning from the trion resonance 
(see Fig.~\ref{level_scheme}) and $\alpha$ gives the
relative phase between two lasers with real Rabi frequencies $\Omega_1(t)$ and $\Omega_2(t)$.  
In the rotating frame given by 
$e^{\mp i\omega_Z t/2}\left|\pm x\right\rangle$, 
$e^{-i(E_t-\Delta)t}\left|t_\uparrow\right\rangle$, the Hamiltonian
is 
\begin{eqnarray}
&&H=\Delta\left|t_\uparrow\right\rangle\left\langle t_\uparrow\right|\nonumber\\
&&+
\tfrac{1}{\sqrt{2}}\left(\Omega_1(t)e^{i\alpha}+\Omega_2(t)e^{i\omega_Zt}\right)
\left|+x\right\rangle\left\langle t_\uparrow\right|+h.c.\nonumber\\
&&+\tfrac{1}{\sqrt{2}}\left(\Omega_1(t)e^{i\alpha-i\omega_Z t}+\Omega_2(t)\right)
\left|-x\right\rangle\left\langle t_\uparrow\right|+h.c.
\label{Hrot}
\end{eqnarray}
In the case $\left|\Omega_{1,2}(t)\right|\ll \omega_Z$, 
the fast oscillating terms can be neglected and the Hamiltonian becomes
\begin{eqnarray*}
H&=&\Delta\left|t_\uparrow\right\rangle\left\langle t_\uparrow\right|+
\tfrac{1}{\sqrt{2}}\Omega_1(t)e^{i\alpha}\left|+x\right\rangle\left\langle t_\uparrow\right|+h.c.\\
&&+\tfrac{1}{\sqrt{2}}\Omega_2(t)(
\left|-x\right\rangle\left\langle t_\uparrow\right|+\left|t_\uparrow\right\rangle\left\langle -x\right|).
\end{eqnarray*}
This Hamiltonian describes the $\Lambda$ system presented 
in Fig.~\ref{level_scheme}. If the temporal profiles of the laser pulses for $\Omega_1(t)$ and
$\Omega_2(t)$ are of rectangular shape
and of the same width, the time dependance of the Rabi frequencies can be omitted.
Assuming $\left|\Omega_{1/2}\right|\ll\Delta$, we obtain
the effective Hamiltonian by the Schrieffer-Wolff transformation \cite{imamoglu_prl1999,
hildmann_prb2011},
\begin{equation*}
H_{{\rm eff}}\approx\begin{pmatrix}
-\frac{\Omega_1^2}{2\Delta} & -e^{i\alpha}
\frac{\Omega_1\Omega_2}{2\Delta}& 0\\
 -e^{-i\alpha}\frac{\Omega_1\Omega_2}{2\Delta}&-\frac{\Omega_2^2}{2\Delta}&0\\
0&0&\Delta+\frac{\Omega_1^2+\Omega_2^2}{2\Delta} \end{pmatrix}.
\end{equation*}
The electron spin states are decoupled from the trion state 
and the effective single spin Hamiltonian reads,
\begin{equation}
H_{{\rm eff}}=\frac{\Omega_2^2-\Omega_1^2}{4\Delta}\,\sigma_z-
\cos\alpha\frac{\Omega_1\Omega_2}{2\Delta}\,\sigma_x
+\sin\alpha\frac{\Omega_1\Omega_2}{2\Delta}\,\sigma_y,
\label{Hspin}
\end{equation}
where $\sigma_i$ are the Pauli matrices in the basis of the states $\left|\pm x\right\rangle$. 
The time evolution operator for the Hamiltonian (\ref{Hspin}) can be 
represented as 
$$U(t)=\exp\left(-i\omega t\, {\hat n}\cdot\vec{\sigma}\right),
$$
where 
$$\omega=\frac{\Omega_1^2+\Omega_2^2}{4\Delta},
$$
$\vec{\sigma}$ is the vector of Pauli matrices and the components of the unit vector $\hat{n}$ are given by
\begin{eqnarray*}
n_x&=&-\cos\alpha\,\frac{2\,\Omega_1\Omega_2}{\Omega_1^2+\Omega_2^2}\equiv-\cos\alpha\,n_\perp,\\
n_y&=&\sin\alpha\,\frac{2\,\Omega_1\Omega_2}{\Omega_1^2+\Omega_2^2}\equiv\sin\alpha\,n_\perp,\\
n_z&=&\frac{\Omega_2^2-\Omega_1^2}{\Omega_1^2+\Omega_2^2}.
\end{eqnarray*}
\section{Hyperfine coupling}
To study the influence of nuclear spins on the performance of the spin rotations, 
we add the hyperfine interaction of an electron confined in a quantum dot to the 
Hamiltonian (\ref{H}), which is given by the contact Fermi interaction \cite{coish_statsolidi2009},
\begin{equation}
 H_{\rm hf}=\mathbf{S}\cdot\mathbf{h}=\mathbf{S}\cdot\sum_{k=0}^{N}A_k\mathbf{I}_k,
\end{equation}
where $\mathbf{S}$ is the electron spin operator and $\mathbf{h}$ is the so called Overhauser field, 
the effective nuclear spin field, $\mathbf{I}_k$ are the nuclear spin operators, $A_k$ is 
the hyperfine coupling strength of a nuclear spin at $k$th lattice site, and $N$ is the
number of nuclear spins interacting with the electron.
According to the central limit theorem, the expectation
value of the Overhauser field underlies a Gaussian distribution with average value at zero and 
with the standard deviation $\sigma=A/\sqrt{N}$, where $A$ is the average hyperfine constant.
For our calculations we used $A=90\,\mu$eV
and $N=10^5$ \cite{cerletti_nano2005}.\\
\indent
The system Hamiltonian together with the hyperfine interaction is
\begin{equation}
 H^h=\begin{pmatrix}\omega_Z/2+h_x/2 & (h_z+i h_y)/2& g^*(t)/\sqrt{2}\\
(h_z-i h_y)/2&-\omega_Z/2-h_x/2&g^*(t)/\sqrt{2}\\
 g(t)/\sqrt{2}&=
g(t)/\sqrt{2} & E_t \end{pmatrix},
\label{Hh}
\end{equation}
where $h_i$ ($i=x,y,z$) are the components of the Overhauser field, 
which is considered here to be a fluctuating effective magnetic field.
If we express the Hamiltonian Eq.(\ref{Hh}) in the same rotating frame as for
Hamiltonian (\ref{Hrot}), it becomes
\begin{widetext}
\begin{equation}
H^h=
\begin{pmatrix} h_x/2 & (h_z+i h_y)e^{i\omega_z t}/2
& (\Omega_1(t)e^{i\alpha}+\Omega_2(t)e^{i\omega_Zt})/\sqrt{2}\\
(h_z-i h_y)e^{-i\omega_z t}/2 &-h_x/2&(\Omega_1(t)e^{i\alpha-i\omega_Z t}+\Omega_2(t))/\sqrt{2}\\
(\Omega_1(t)e^{-i\alpha}+\Omega_2(t)e^{-i\omega_Zt})/\sqrt{2} &
 (\Omega_1(t)e^{-i\alpha+i\omega_Zt}+\Omega_2(t))/\sqrt{2} & \Delta \end{pmatrix}.
\label{Hhrot}
\end{equation}
\end{widetext}
We can neglect again the fast oscillating terms under the assumption 
$\left|\Omega_{1,2}(t)\right|, \sigma\ll\omega_Z$. In this way, the transverse terms from the 
hyperfine coupling are excluded from our calculations and the Hamiltonian in the rotating frame 
with included Overhauser field becomes
\begin{equation}
H^h\simeq\begin{pmatrix}h_x/2 & 0& \Omega_1(t)e^{i\alpha}/\sqrt{2}\\
 0&-h_x/2&\Omega_2(t)/\sqrt{2}\\
\Omega_1(t)e^{-i\alpha}/\sqrt{2}&\Omega_2(t)/\sqrt{2}& \Delta \end{pmatrix}.
\end{equation}
Choosing again the laser profiles of rectangular shape and the same width for $\Omega_{1/2}(t)$ and assuming
that the Overhauser field is static, we can render the Hamiltonian $H^h$ time independent.
The system undergoes the dynamics given by $H^h$ only during the laser pulse.
Applying again the Schrieffer-Wolff transformation and treating the nuclear field as a random field
we obtain an effective Hamiltonian.
Also here, the electron spin states are decoupled
from the trion state and we can work only 
on the electron spin states subspace.
For the single spin Hamiltonian we find (up to a constant)
\begin{eqnarray}
&& H^h_{\rm eff}\approx\frac{1}{2}\left(h_x-\frac{\Omega_1^2(t)}{2\Delta-h_x}+
\frac{\Omega_2^2(t)}{2\Delta+h_x}\right)\sigma_z
\nonumber\\
&&-\cos\alpha\,\frac{2\Delta\Omega_1(t)\Omega_2(t)}{4\Delta^2-h_x^2}\sigma_x
+\sin\alpha\frac{2\Delta\Omega_1(t)\Omega_2(t)}{4\Delta^2-h_x^2}\,\sigma_y. \label{Hhspin}
\end{eqnarray}
The unitary time evolution operator of this Hamiltonian can be represented as
\begin{equation}
U_h(t)=\exp\left(-i \omega(h)t\,\hat{n}(h)\cdot\vec{\sigma}\right),
\end{equation}
where 
\begin{eqnarray*}
&&\omega(h)=\frac{1}{2(4\Delta^2-h_x^2)}\left(16\Delta^2\Omega_1^2\Omega_2^2+
\right.\\
&&\left.\left(2\Delta(\Omega_2^2-\Omega_1^2)-h_x(\Omega_2^2+\Omega_1^2)
+h_x(4\Delta^2-h_x^2)\right)^2\right)^{1/2}.
\end{eqnarray*}
The components of the unit vector $\hat{n}(h)$ are
\begin{eqnarray*}
n_x(h)&=&-\cos\alpha\,n_\perp(h),\\
n_y(h)&=&\sin\alpha\,n_\perp(h),\\
n_z(h)&=&\frac{2\Delta(\Omega_2^2-\Omega_1^2)-h_x(\Omega_2^2+\Omega_1^2)+h_x(4\Delta^2-h_x^2)}
{2(4\Delta^2-h_x^2)\omega(h)},
\end{eqnarray*}
with
\begin{equation*}
n_\perp(h)\equiv
\frac{2\Omega_1\Omega_2\Delta}{(4\Delta^2-h_x^2)\omega(h)}.
\end{equation*}
%
%and
% 
%\begin{equation}
%n_\parallel(h)\equiv
%\frac{2\Delta(\Omega_2^2-\Omega_1^2)-h_x(\Omega_2^2+\Omega_1^2)+h_x(4\Delta^2-h_x^2)}
%{2(4\Delta^2-h_x^2)\omega(h)}.
%\end{equation}
%
\section{Fidelity}
The deviation in the optical rotations of the electron spin state due to the coupling to nuclear
spins is studied by calculating the fidelity of the time evolution.
The fidelity for two unitary operators 
averaged over all possible initial states on which the operators are acting
is given by \cite{petersen_pra2007, ghosh_pra2010}
\begin{equation}
\mathcal{F}=\frac{n+|{\rm Tr}[U^{\dagger}_{\rm ideal}U_{\rm actual}]|^2}{n(n+1)},
\end{equation}
where $n$ is the dimension of the Hilbert space, $U_{\rm ideal}$ represents the ideal operator and
$U_{\rm actual}$ is the actual operator.
The fidelity of single spin rotation ($n=2$)
in the presence of nuclear spins is given by
$$\mathcal{F}=\frac{1}{3}+\frac{1}{6}\left|\mbox{Tr}[U(t)^\dagger U_h(t)]\right|^2.
$$
\begin{figure}[ht!]
\centering\includegraphics[width =0.45\textwidth]{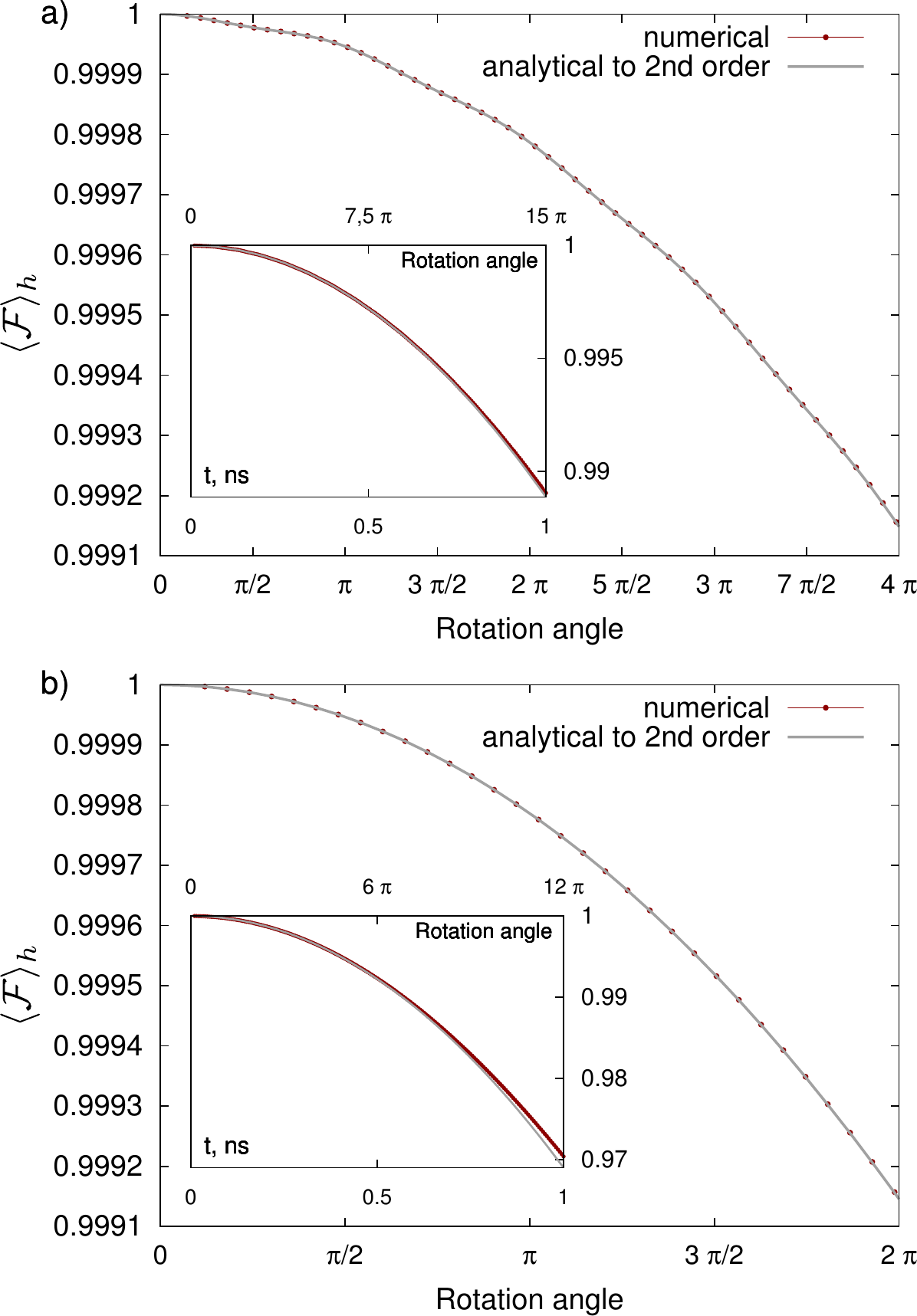}
\caption{a) Fidelity for a generic spin rotation averaged numerically over Overhauser field
distribution and its the average value to second order of the Overhauser field. Here $\Omega_1=1$ meV, 
$\Omega_2 =$0.5 meV and $\Delta=$ 10 meV.
Inset:
the average fidelity for a generic rotation on a longer time scale.
The spin is rotated by around $15\pi$ after 1 ns. 
b) Fidelity for a spin rotation around the $x$ axis averaged numerically over Overhauser field
distribution and its the average value to second order of the Overhauser field. Here $\Omega_1=1$ meV, 
$\Omega_2 =$0 and $\Delta=$ 10 meV.
Inset:
the average fidelity for a rotation around $x$ axis on a longer time scale (here 1 ns corresponds
to a rotation angle
of around $12\pi$).}
\label{f1}
\end{figure}
\begin{figure}[ht!]
\centering\includegraphics[width =0.45\textwidth]{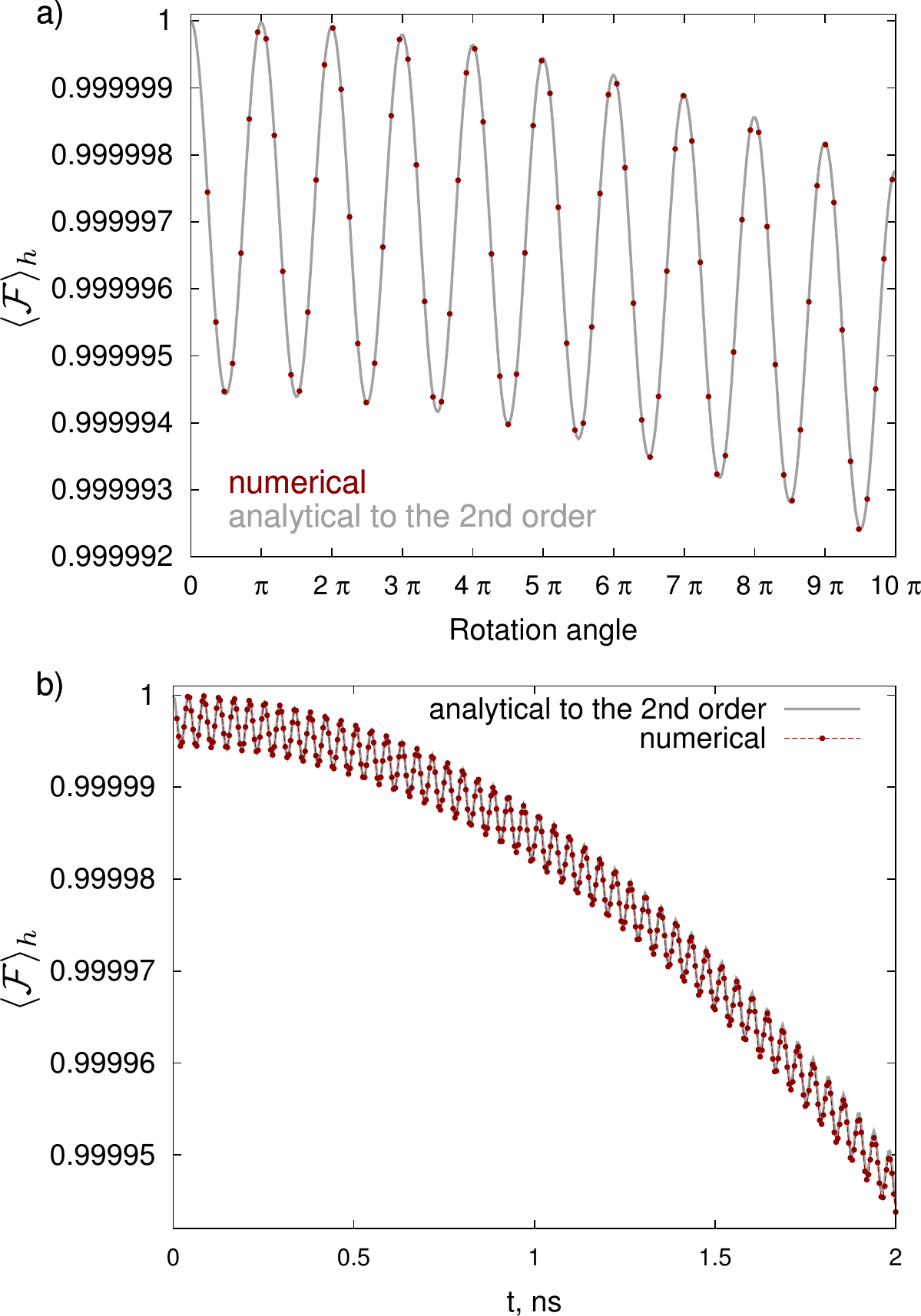}
\caption{Fidelity for rotations around an axis nearly in the $y$-$z$ plane averaged over nuclear spins 
numerically and to second order of hyperfine interaction: a) for short times and b)
for longer duration of the laser pulses. Used parameters: $\Omega_1$ = 1 meV, $\Omega_2$ = 0.98 meV
and $\Delta =$ 10 meV.}
\label{f2}
\end{figure}
The trace of the product of the perfect time evolution operator and 
the time evolution operator with Overhauser effect is given by
\begin{eqnarray*} 
\mbox{Tr}[U(t)^\dagger U_h(t)]&=&2\left(\cos\omega t\cos\omega(h)t+\right.\\
&&\left.\sin\omega t\sin\omega(h) t[n_\perp n_\perp(h)+n_z n_z(h)]\right).
\end{eqnarray*}
To obtain the average fidelity, we need to average the following expression 
analytically to a particular order
or numerically over the Overhauser field distribution,
\begin{eqnarray}
\mathcal{F}&=&\frac{1}{3}+\frac{2}{3}\left(\cos\omega t\cos\omega(h)t+\right.\label{fid}\\
&&\left.\sin\omega t\sin\omega(h) t[n_\perp n_\perp(h)+n_z n_z(h)]\right)^2.\nonumber
\end{eqnarray}
The average fidelity to the second order of the Overhauser field is given by
\begin{eqnarray}
\left\langle \mathcal{F}\right\rangle_h&=&1-
\frac{(\Omega_1^2+\Omega_2^2-4\Delta^2)^2(\Omega_1^2-\Omega_2^2)^2}
{96\Delta^4(\Omega_1^2+\Omega_2^2)^2}t^2\sigma^2\label{fid2}\\
&&-\frac{2(\Omega_1^2+\Omega_2^2-4\Delta^2)^2\Omega_1^2\Omega_2^2}
{3\Delta^2(\Omega_1^2+\Omega_2^2)^4}\,\sigma^2\,\sin^2\omega t\, +
\mathcal{O}(\sigma^4),\nonumber
\end{eqnarray}
where $\sigma=\langle h_x^2\rangle$ is the standard deviation of the Overhauser field distribution. 
The numerically obtained average fidelity is presented in comparison to the
analytical result to second order of hyperfine interaction in 
Fig.~\ref{f1} and \ref{f2}.\\ 
\begin{figure}[ht!]
\centering\includegraphics[width =0.45\textwidth]{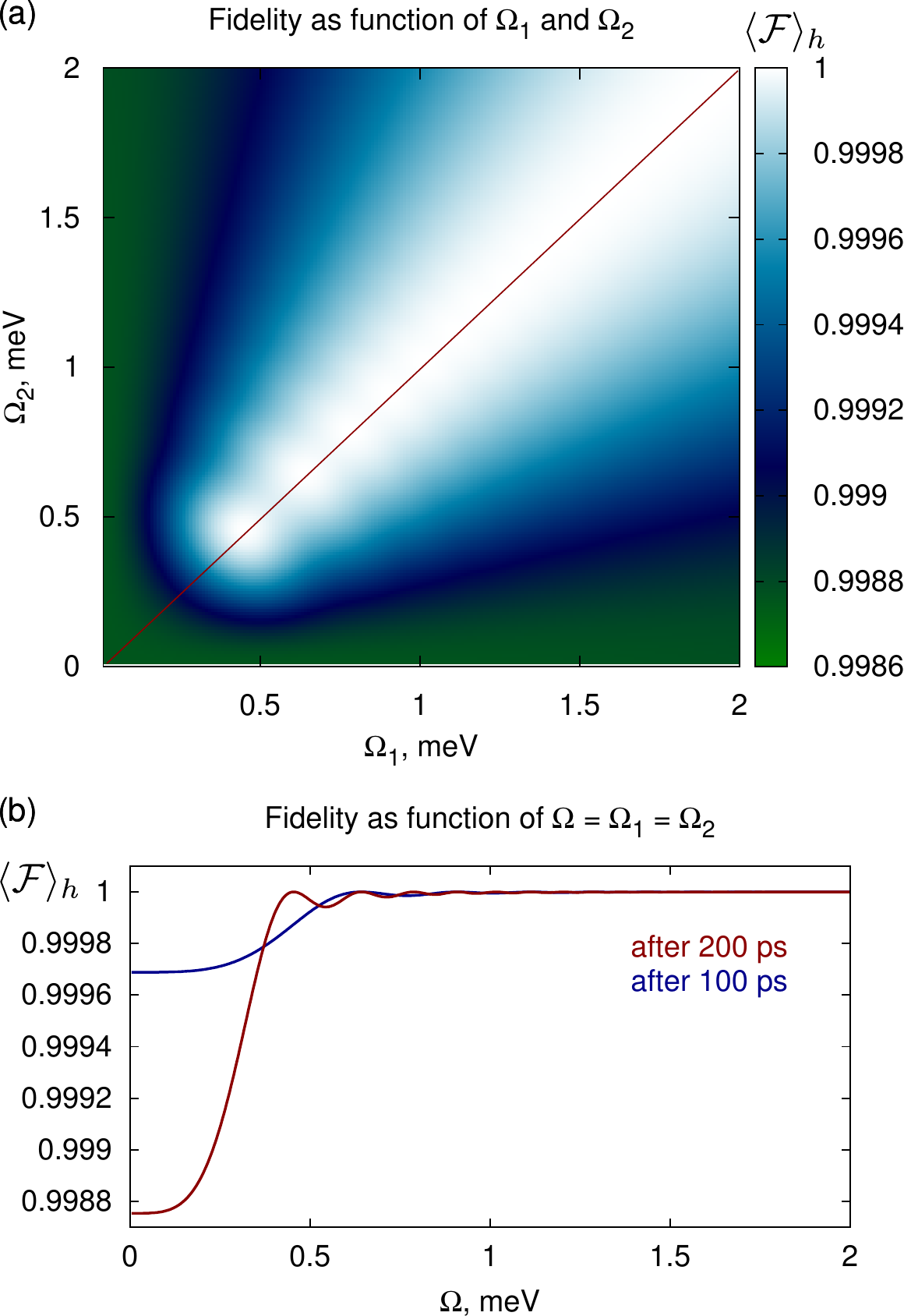}
\caption{a) Average fidelity to second order of hyperfine interaction for $\Delta=10$ meV as function
of Rabi frequencies after $t=$ 200 meV. b) The special case of the average fidelity,
when Rabi frequencies are equal:
$\Omega_1=\Omega_2\equiv\Omega$ as function of $\Omega$ at $\Delta= 10$ meV for different interaction times:
100 ps (blue) and 200 ps (red).}
\label{fpar}
\end{figure}
\indent
Fig.~\ref{f1}(a) shows the average fidelity for a rotation around a generic axis, which
is not parallel or perpendicular to the characteristic axes. In this case $\Omega_1\neq\Omega_2$ and
$\Omega_{1,2}\neq0$
as it can be seen in Eq. (\ref{Hspin}) and (\ref{Hhspin}). For the parameters $\Omega_1$ = 1 meV,
$\Omega_2$ = 0.5 meV and $\Delta$ =10 meV the duration of a $\pi/2$ rotation is 35 ps and 
the average fidelity is 0.999978. The average fidelities obtained both numerically and using
second-order perturbation theory, Eq.~\eqref{fid2}, agree for short interaction times and remain in good
agreement up to the nanosecond scale.\\
\begin{figure}[ht!]
\centering\includegraphics[width =0.45\textwidth]{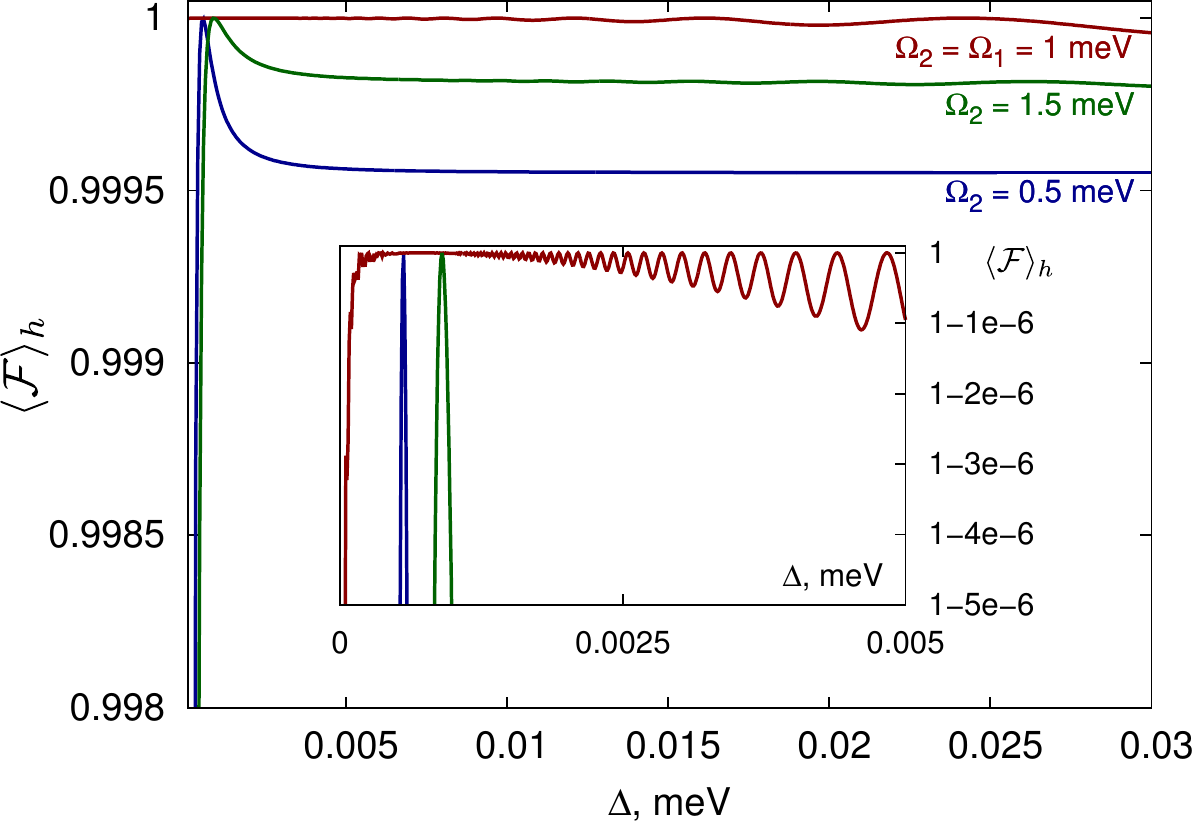}
\caption{Average fidelity after 200 ps interaction time as function of the laser detuning for different 
Rabi frequencies: blue plot gives the average fidelity for $\Omega_1=0.5$ meV, $\Omega_2=1$ meV, green for
$\Omega_1=0.5$ meV, $\Omega_2=1$, and red plot is the special case of $\Omega_1=\Omega_2$. Inset: the oscillations
of the fidelity $\propto\sin^2\omega t$ to the unity in the case of equal Rabi frequencies. }
\label{fdet}
\end{figure}
\indent
The rotations around the axis along the magnetic field ($x$ axis here) are obtained by setting
$\Omega_1$ or $\Omega_2$ equal to zero. The average fidelity for such a rotation is shown in
Fig.~\ref{f1}(b) with 
$\Omega_1 = $ 1 meV, $\Omega_2$ = 0 and $\Delta$ = 10 meV. For these parameters a $\pi/2$ rotation 
lasts around 41 ps and the average error for the rotation is $1-\langle \mathcal{F}\rangle_h=5.3\cdot 10^{-5}$.
The reduction of the average fidelity
in this case is anticipated, since the rotation frequency $\omega$
is a quadratic function of both Rabi frequencies and the reduction of these frequencies leads to smaller 
$\omega$ and thus to smaller fidelities. The average fidelity obtained 
analytically to second order of the Overhauser field coincides with the numerically averaged fidelity
for few full rotations around the $x$ axis and reproduces it on the nanosecond scale
(Fig.~\ref{f1}(b)).  \\
\indent
The rotations around an axis lying in the $y-z$ plane (corresponding to the $\sigma_x$ and
$\sigma_y$ terms in Eq.~\eqref{Hspin} and \eqref{Hhspin}) can be obtained by applying
pulses with $\Omega_1=\Omega_2$.
Furthermore, the rotation axis is specified by the phase $\alpha$ as shown in Eq.~\eqref{Hspin}.
However setting $\Omega_1=\Omega_2$ does not result in a rotation
around an axis lying in the $y-z$ plane, if the electron spin interacts with the nuclear spins.
The axis is rotated out of the $y-z$ plane because of the Overhauser field  as can be seen in
Eq.~\eqref{Hhspin}.
The average fidelity of a single spin rotation around an axis,
that is defined by 
$\Omega_1=1$ meV and $\Omega_2=0.98$ meV is shown in Fig.~\ref{f2}.
The fidelity averaged over the nuclear spin distribution
numerically to an arbitrary order and averaged analytically to the second order
of the Overhauser field $h$ is enhanced compared 
to the average fidelities in other cases presented in Fig.~\ref{f1}.
The duration of a $\pi/2$ rotation reduces to 20 ps, while the
average error for such rotation decreases to $1-\langle \mathcal{F}\rangle_h=3\cdot 10^{-6}$.
this strong improvement in fidelity cannot be explained only by increase of the 
interaction energy with the laser light $\omega$.
The increased average fidelity in the case, when $\Omega_1$ is close to $\Omega_2$ and vice versa
can be attributed to the interplay of different contributions leading to a reduction of the average
fidelity. As it can be seen in Eq.~\eqref{fid2}
the second term reduces the fidelity as $\propto t^2$ and the third term as $\propto \sin^2\omega t$.
When both Rabi frequencies $\Omega_1$ and $\Omega_2$ are of roughly the same value,
the fidelity reducing term $\propto t^2$ becomes less
relevant compared to the term
$\propto\sin^2 \omega t$. This can be observed in Fig.~\ref{f2}(a), where the average fidelity exhibits an
oscillatory
behavior for few full rotations around the given axis.
The fidelity oscillations become dominated by the $\propto t^2$ decay for longer interaction times 
(Fig.~\ref{f2}(b)). \\
\indent
The average fidelity depends not only on time (duration of the laser pulse),
it also depends on the detuning and the Rabi frequencies, since they affect
the interaction strength and rotation axis of the applied pulse.
The density plots in Fig.~\ref{fpar} show the dependence of the average
fidelity on the Rabi frequencies and the laser detuning. The calculations were done 
for a fixed time duration of 200 ps, which corresponds to a different rotation angle
depending on the two Rabi frequencies. Fig.~\ref{fpar}(a) shows
the dependence of the average fidelity on the two Rabi frequencies. The average fidelity increases
as the Rabi frequencies grow, since it increases the interaction energy $\omega$ and
shortens the time needed to perform certain rotations. What is remarkable here is
the strong enhancement of the
fidelity (up to unity) in the region, where $\Omega_1=\Omega_2$. 
This means that the average fidelity of a single spin
rotation depends on the rotation axis in addition to the rotation frequency.
The rotations around axes perpendicular
to the applied magnetic field are the least sensitive to the nuclear spin effects.
The cut of the density plot in 
Fig.~\ref{fpar}(a) for $\Omega_1=\Omega_2$ is shown in Fig.~\ref{fpar}(b). As it can be seen in both
Fig.~\ref{fpar}(a) and \ref{fpar}(b) for small $\Omega$ the average fidelity has a constant value and then
increases and oscillates. From the formula \eqref{fid2} we have for the average fidelity at the case 
$\Omega_1=\Omega_2\equiv\Omega$ and under assumption, that $\Omega\ll\Delta$:
$$\langle\mathcal{F}\rangle_h\approx 1-\frac{2}{3}\frac{\Delta^2}{\Omega^4}
\sigma^2\sin^2\frac{\Omega^2 t}{2\Delta}.
$$  
For $\Omega\geq\sqrt{2\Delta/t}$ the fidelity oscillates with decreasing period and
for $\Omega\ll\sqrt{2\Delta/t}$ (for $t=200$ ps and $\Delta=10$ meV this threshold is around 80 $\mu$eV)
is given by $\langle\mathcal{F}\rangle_h\approx1-\sigma^2t^2/6$,
which for $t=200$ ps is $\langle\mathcal{F}\rangle_h\approx0.99875$.
The last expression describes the average fidelity to second
order of hyperfine interaction for an electron spin interacting just
with a static magnetic field and nuclear spins.
As it is shown in Fig.~\ref{fpar}(b), this fidelity increases for shorter interaction times.\\
\indent
The behavior of the average fidelity for spin rotations at $\Omega_1=\Omega_2$ has a special character:
it reaches unity when $\sin\omega t=0$. This phenomenon is due to the overlap of two effects
Without nuclear spins
this situation corresponds to a rotation around an axis in the $y-z$ plane,
but with the hyperfine interaction 
there is an additional fluctuation of the rotation axis perpendicular to the $y-z$ plane. In the case of 
$\Omega_1=\Omega_2$ these fluctuations average to zero. This leads to the result
that only the off-diagonal
elements of the time evolution operator are altered by the hyperfine interaction. But exactly this effect cannot
be captured by the fidelity, when $\sin\omega t=0$, because in this case the ideal time evolution
operator is the identity operator. Consequently, the trace of the product of the time evolution operators, 
the ideal one and 
one with hyperfine interaction, does not contain the off-diagonal terms of the affected time evolution operator,
which results in a perfect fidelity.
This can also be seen in the dependence of the fidelity on the laser detuning in Fig. \ref{fdet}.  
For differing Rabi frequencies the average fidelity reaches unity only once at 
$\Delta=\sqrt{\Omega_1^2+\Omega_2^2}/2$, which corresponds to the situation
when the optical Stark shift compensates
the effective magnetic field induced by nuclei, and then decays to a value given by 
$1-\sigma^2t^2(\Omega^2_1-\Omega_2^2)^2/(6(\Omega_1^2+\Omega_2^2)^2)$. For equal Rabi frequencies
the fidelity oscillates with increasing period and reaches unity again for $\sin\omega t=0$.
\section{Conclusions}
We studied the nuclear spin effect on the performance of Raman
assisted optical transitions of an electron spin in a semiconductor quantum dot. 
It was shown that the average rotation fidelities obtained using second-order perturbation theory
are in good agreement up to nanoseconds time scale
with the numerically averaged fidelities.
The average fidelities were calculated for different rotation axes using both approaches.
In the framework of the formalism used for describing the interaction of the single electron spin
with the laser light
the average rotation fidelities differ strongly for different rotation axes.
While the rotations around axes in the $y-z$ plane, perpendicular to the
applied magnetic field suffer at least under interaction with nuclear spins,
the rotations around the x-axis, which is parallel to the external magnetic field, are the most
effected by the hyperfine interaction.
\section{Acknowledgments}
We gratefully acknowledge funding from the DFG within SFB 767 and from BMBF under the program QuaHL-Rep.

\end{document}